# Ribonucleic acid (RNA) virus and coronavirus in *Google Dataset Search*: their scope and epidemiological correlation

Manuel Blázquez-Ochando
Complutense University of Madrid
https://orcid.org/0000-0002-4108-7531

Juan-José Prieto-Gutiérrez
Complutense University of Madrid
https://orcid.org/0000-0002-1730-8621

**Abstract**
This paper presents an analysis of the publication of datasets collected via *Google Dataset Search*, specialized in families of RNA viruses, whose terminology was obtained from the *National Cancer Institute* (NCI) thesaurus developed by the US *Department of Health and Human Services*. The objective is to determine the scope and reuse capacity of the available data, determine the number of datasets and their free access, the proportion in reusable download formats, the main providers, their publication chronology, and to verify their scientific provenance. On the other hand, we also define possible relationships between the publication of datasets and the main pandemics that have occurred during the last 10 years. The results obtained highlight that only 52% of the datasets are related to scientific research, while an even smaller fraction (15%) are reusable. There is also an upward trend in the publication of datasets, especially related to the impact of the main epidemics, as clearly confirmed for the Ebola virus, Zika, SARS-CoV, H1N1, H1N5, and especially the SARS-CoV-2 coronavirus. Finally, it is observed that the search engine has not yet implemented adequate methods for filtering and monitoring the datasets. These results reveal some of the difficulties facing open science in the dataset field.

## 1. Introduction

The consultation of data sets and articles, located in the repositories, has become a common practice and a central role in the investigation of people or work groups (Marcial; Hemminger, 2010) for making well-founded decisions (Hernández- Pérez, 2016). For example, the mean annual size of the data set in the Miccai articles (Medical Image Computing and Computer-Assisted Intervention) has grown approximately 3 to 10 times between 2011 and 2018 (Landau; Kiryati, 2019), which comes to confirm this approach and reveals the paradigm shift towards an open science.

And that is why every day there is greater awareness of the need to share data and materials derived from scientific research to reproduce analyzes, compare them and pose new questions, (Nosek et al., 2015) although it is true that certain concerns are produced subject to the confidentiality, governance, possible misuse of institutional and business data (Howe et al., 2018). Despite the fact that the disadvantages are less than the benefits, according to a survey answered by 800 researchers, which concluded that less than 8% considered the possible negative consequences of data exchange (Mello; Lieou; Goodman, 2018).

The latest open data works (Corrales-Garay; Ortiz-de-Urbina-Criado; Mora-Valentín, 2019) show a more than probable increase in its use, due to:
change of behavior in scientific research, which is now also based on massive data analysis or big-data (Saheb; Izadi, 2019); control of data protection laws (Polonetsky; Tene; Finch, 2016); increased transparency (Weston et al., 2019); Requirements of funding agencies, which require supporting scientific conclusions on proven and recognizable data in shared datasets.

This can be verified in the Plan S of the European Commission (Science Europe, 2019) which urges, from January 1, 2021, the publication in golden open access journals or in repositories and "related platforms" that publish Editorial PDF, being also a great opportunity for magazines to carry out a full digital transformation (López-Borrull et al., 2020).

Recently, open access scientific preprints are becoming a fundamental source of information to face transcendental issues, such as the health crisis produced by SARS-CoV-2 (Johansson; Saderi, 2020), in which researchers from all over the world are uniting their efforts, knowledge and databases to identification of infected patients through the symptoms of fever and its incidence pattern (Haleem et al., 2020);
-space-time prediction of the speed and magnitude of virus transmission (Zhou et al., 2020); simulation of protein folding for targeted therapies (Chen et al., 2020); predicting the progress of Covid-19 disease through radiological images (Chen; Lerman; Ferrara, 2020); to obtain effective treatments and vaccines (Le-Guillou, 2020).

The purpose of this work is the analysis of the results obtained in the Google dataset search engine, in reference to virus families, being necessary to answer the following questions:
Is Google Dataset Search an open data search engine, suitable for research?
What is the ratio of open and reusable datasets for preset virus queries?
Which RNA viruses have the greatest number of research datasets?
What is the evolution of the datasets depending on the virus?
Is there a correlation between the chronology of the epidemics and the publication of datasets and scientific articles?
What documentary difficulties are found in open scientific datasets?
How can the obtained datasets be reused?

It will be observed that the research questions are oriented in a double aspect of Information-Documentation and correlative of viruses.

This is so because the number of datasets and their evolution is related, among other issues, to viral diseases, as explained in the research.

The search engine Google Dataset Search is a query utility specialized in data sets, which collects information from scientific, commercial and governmental repositories of a very diverse nature, as will be explained later. The range of this tool is not comparable with other search engines and aggregators, not even with the data providers that it covers in a single corpus.

In addition, Google collects datasets for its search engine that comply with the schema.org metadata standard (Brickley; Burgess; Noy, 2019), more specifically that referring to datasets, as explained in its approach to dataset discovery (Google, 2020). This has allowed the creation of a massive compilation instrument that undoubtedly facilitates scientific research. However, it also raises questions about its suitability, as will be explained later. This search engine entered service on September 5, 2018, but was not definitively open until January 2020, at which time it has coincided with the Coronavirus crisis, so it will be able to demonstrate its versatility, advantages and shortcomings, in a situation of urgency and necessity. In this research, additionally, an attempt will be made to observe the limitations of the search engine, as well as the correctness of the information presented, its possibilities for reuse and relevance, carrying out specialized searches in virology.

## 2. Methodology

The methodology used in this work is aimed at preparing a sample of queries from a controlled vocabulary, which allows the Google Dataset Search to be interrogated. This method is commonly used in search engine evaluation works (Lewandowski, 2015), in order to observe the relevance and pertinence of the results, apart from other evaluation methods (Hawking et al., 2001). For this, the selection of a vocabulary sample is required, from a documentary language, that is, a controlled and normalized vocabulary, in order to ensure the repeatability and reproducibility of the tests on the search engine, according to (Broder, 2002). This is also representative of the TREC consultation and evaluation methodology (Hawking et al., 1999). Each word or noun phrase chosen serves to compose a query, for which a query formula is designed whose complexity will vary depending on the purpose of the evaluation. In this investigation, as will be explained later, no complex equations are required, just searching for literal terms in the search variable by default. This allows creating a search url that can be executed manually or automatically, in order to collect the results that the search engine returns. In this research, the following are observed: the quantitative values of the total number of results of the query; the proportion of the results according to the type of access and format; the publication and registration chronology of the datasets; the number of scientific articles linked to the retrieved datasets; the provenance of the data sets (these are the main data providers); the proportion of reusable datasets for research and documentation tasks; the relevance of the results according to each query raised (that is, the adaptation of the results to the term of the vocabulary used) or what is the same, the analysis of the content offered by the search engine for each query (King et al., 2007).

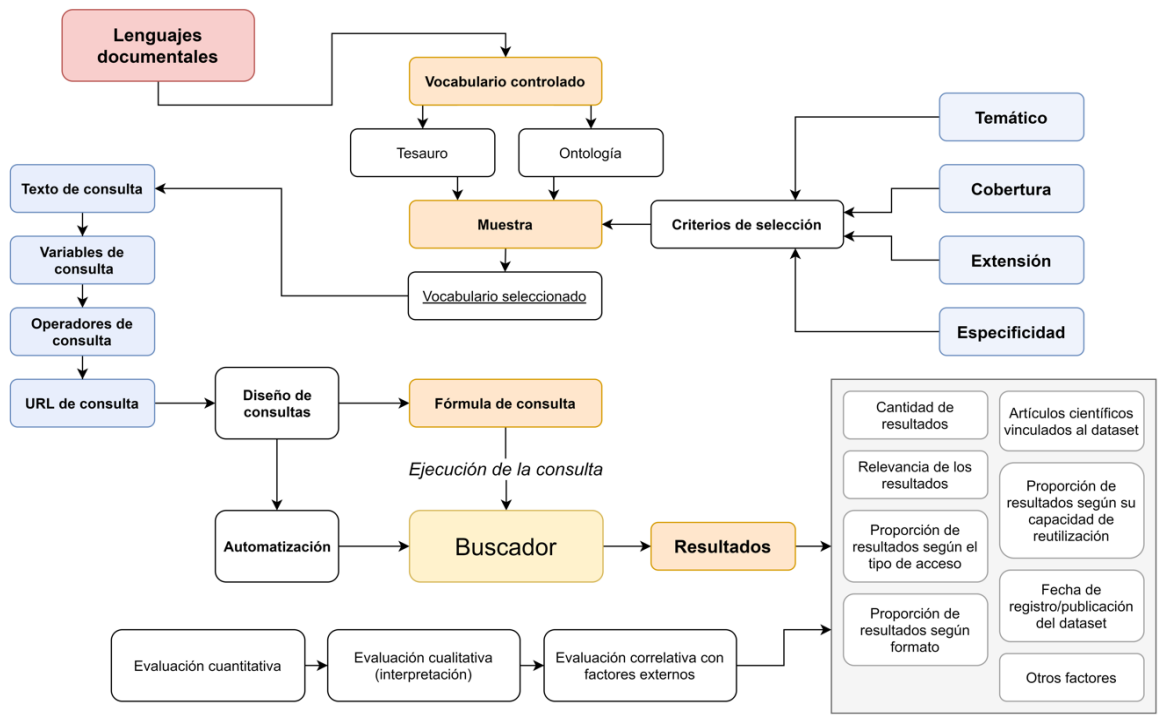

Figure 1. Methodological model for evaluating search engines, used in the research.

Next, the phases of the method are studied in depth: a) Selection of specialized terminology; b) Configuration and execution of queries in Google Dataset Search; c) Analysis of results.

- For the selection of terminology specialized in virus families, the thesaurus of the NCI (National Cancer Institute) has been chosen, which is used extensively by the American NIH (National Institute of Health). Next, the section of the thesaurus has been selected, which contains the terminological references of Viruses (Code C14283), which include DNA viruses, retroviruses, RNA viruses and other groups of viruses. For this research, RNA viruses have been chosen because they are found among them, those related to the Coronaviridae family, which have caused the international health alert in 2020. In this way, the number of datasets recovered from SARS- can also be quantitatively compared. CoV-2, with respect to the rest of viruses of the same classification. In sum, a total of 22 families of RNA-type viruses and a total of 70 terms have been obtained, which can be verified in Table 1.

Table 1. Terms obtained from the NCI thesaurus, which have been used to make the queries in Google Dataset Search

| Virus families (Q1) | Virus (Q2) |
|---|---|
| Reoviridae | Colorado tick fever virus, Orbivirum, Rotavirus |
| Arenavirus | Lymphocytic choriomeningitis virus, Tacaribe virus |
| Bunyaviridae | Hantavirus, Nairovirus, Orthobunyavirus, Phlebovirus |
| Filoviridae | Ebola virus, Marburgvirus |
| Influenza | Avian influenza, H5N2 avian influenza, Influenza H1N1, Influenza H5N1 |
| Avulavirus | Newcastle disease virus |
| Henipavirus | Hendra virus, Nipah virus |
| Morbillivirus | Measles morbillivirus |
| Paramyxovirus | Human parainfluenza |
| Respirovirus | Human parainfluenza virus 1, Human parainfluenza virus 3 |
| Rubulavirus | Human parainfluenza virus 2, Human parainfluenza virus 4, Mumps virus |
| Pneumovirinae | Human respiratory syncytial virus, Metapneumovirus, Pneumovirus |
| Rhabdoviridae | Rabies virus |
| Astroviridae | Astroviridae |
| Hepatitis E | Hepatitis E |
| Norovirus | Norovirus genogroup, Norwalk virus |
| Sapovirus | Sapporo Virus |
| Coronavirus | Porcine epidemic diarrhea virus, SARS Coronavirus, SARS Coronavirus 2, Covid-19, Covid-20, Covid-21, Covid-22, Covid-23, Deltacoronavirus, Gammacoronavirus |
| Flavivirus | Dengue virus, Powassan virus, Tick-Borne encephalitis virus, West Nile virus, Yellow fever virus, Zika virus, Hepatitis C virus, Pegivirus |
| Picornavirus | Aphthovirus, Coxsackie A virus, Coxsackie B virus, Echovirus, Enterovirus D68, Enterovirus D69, Enterovirus D70, Enterovirus D71, Enterovirus D72, Poliovirus, Hepatitis A virus, Rhinovirus |
| Togaviridae | Togaviridae |
| Alphavirus | Barmah forest virus, Chikungunya virus, Ross river virus, Rubella virus, Rubivirus |

- Configuration and execution of queries in Google Dataset Search. Two types of queries have been considered. First, searches for virus families, which will be codenamed Q1 queries, and secondly, searches for virus species, which will be called Q2 queries. The main data providers, dataset formats, their reuse, type of access and the chronology of their publication will be obtained for each species of virus. It is relevant to specify that a result given by the search engine can assume one or more datasets, since they can often be found in different export formats. In this work, the number of datasets has been specifically considered, to obtain greater precision in the figures provided. The query mode is based on the direct use of the thesaurus descriptors, to which literal search quotes are added, for example "Porcine Epidemic Diarrhea Virus", in order to obtain the most relevant datasets for each virus in question.

- Analysis of results. The analysis process is intended to compare the data from queries Q1 and Q2. These data will be compared with a Q2 sample, configured with the datasets of the first 5 results of each virus species, for prospective analysis. Possible quantitative and qualitative differences will be observed, their frequency of publication, correlation with the main epidemics and pandemics, evolution in the last 20 years and their comparison with the publication of scientific articles. Additionally, the limitations and problems observed in Google Dataset Search will be pointed out.

## 3. Results

The Q1 category queries, related to virus families, yield a total of 1,375 datasets, of which almost 87% are open access. By filtering the results by specific formats, suitable for the reuse of information, that is, the CSV (comma separated values) and XLS (Excel spreadsheet or similar) format, it is obtained that only 58.59% can be used in databases and statistical analysis programs, since the remaining 41.41% are in image formats, PDF or office type documents.

The results of the Q2 category show proportions similar to those already observed in Q1. 82% of the results correspond to free access datasets and the remaining 18% to subscription datasets. On the other hand, 64% is in usable formats and 36% in image or derived office formats. These preliminary data confirm the rise of open science in this sector.

However, when examining the Q2 sample carefully, very different results are observed, which raise serious doubts regarding the information that Google Dataset Search provides a priori. Of the 3,799 datasets obtained, those of the first 5 most relevant results for each virus species are analyzed in depth, obtaining a sample of 331 datasets. From their scrutiny, it was concluded that only 67 datasets are reusable, being available in standard CSV, XLS, SQL, XML formats. This means that only 21% of the total is usable, which means a disagreement with the information given by the search engine in the Q1 and Q2 queries.

Table 2. Results obtained for the consultations Q1, Q2 and sample of Q2.

| | Q1 | | Q2 | | sample Q2 | |
|---|---|---|---|---|---|---|
| | n. datasets | % | n. datasets | % | n. datasets | % |
| **n. total datasets retrieved** | **1.375** | **100%** | **3.799** | **100%** | **331** | **100%** |
| Distribution according to access | | | | | | |
| Open Access | 1.197 | 86,80% | 2.890 | 76,07% | 317 | 95,77% |
| Subscription or payment | 182 | 13,20% | 641 | 23,93% | 14 | 4,22% |
| Distribution according to reuse | | | | | | |
| Reusable (.csv, .xls, .sql, .xml) | 808 | 58,59% | 2.423 | 63,78% | 71 | 21,45% |
| Not reusable (.pdf, .doc, .ppt, .png, .tif, .jpg) | 571 | 41,41% | 1.376 | 36,22% | 260 | 78,55% |

Source: Google Dataset Search. Collection date: March 8, 2020

Delving into the data obtained, in table 3 it is observed that the number of reusable datasets is reduced to 15%, if the criterion of proven scientific provenance is added. That is, that at least the link between the dataset and a scientific article is verified. It has also been discovered that 37% of the non-reusable datasets had scientific origin, corresponding mostly to images, illustrations and scientific papers, which means that they are not actually data sets. On the other hand, there is also a margin of 6% of the datasets that, being reusable, did not have an accredited scientific origin, due to the lack of attached research. Furthermore, 66% of the published datasets are concentrated in the last 5 years, which indicates that the data update frequency may be low. To these figures, we must add the factor of aggregation and heterogeneity of the data, since, in many similar datasets, the content object, its structure and variables did not coincide, which means that they are not continuous in most cases. Consequently, their aggregation and continuity factor are very low.

Table 3. Chronological distribution of the datasets analyzed in the Q2 sample

| Date | Analyzed datasets | Scientists | Non-scientific | Reusable (.csv, .xls, .sql) | Not reusable (They are not datasets) | Reusable of scientific origin | Not reusable of scientific origin | Reusable without accredited scientific origin |
|---|---|---|---|---|---|---|---|---|
| < 2010 | 87 | 4 | 89 | 1 | 86 | 1 | 3 | 0 |
| 2010 | 1 | 0 | 1 | 0 | 1 | 0 | 0 | 0 |
| 2011 | 1 | 0 | 0 | 1 | 0 | 0 | 0 | 0 |
| 2012 | 10 | 2 | 5 | 0 | 10 | 0 | 2 | 0 |
| 2013 | 12 | 4 | 8 | 0 | 12 | 0 | 4 | 0 |
| 2014 | 19 | 3 | 16 | 2 | 17 | 1 | 2 | 1 |
| 2015 | 34 | 29 | 5 | 13 | 21 | 11 | 18 | 2 |
| 2016 | 55 | 54 | 1 | 11 | 44 | 11 | 43 | 0 |
| 2017 | 16 | 12 | 3 | 5 | 11 | 4 | 8 | 1 |
| 2018 | 28 | 22 | 6 | 9 | 19 | 6 | 16 | 3 |
| 2019 | 40 | 30 | 9 | 10 | 30 | 7 | 23 | 3 |
| 2020 | 28 | 12 | 16 | 19 | 9 | 9 | 3 | 10 |
| Sum | 331 | 172 | 159 | 71 | 260 | 50 | 122 | 20 |
| % | 100% | 52% | 48% | 21% | 79% | 15% | 37% | 6% |

Figure 2 shows the absolute publication frequency of the data sets for each virus. The number of published datasets does not show a linear progression until 2010, at which time there is constant and sustained growth. In 2016, the largest number of data sets related to the virus test were recovered, highlighting:

- Lymphocytic Choriomeningitis or Choriomeningitis virus, produced by rodents (29 results);
- Nipah virus endemic to the area of Malaysia and India (31 results);
- the Rhinovirus or common cold virus (25 results);
- the SARS Coronavirus or SARS-CoV (25 results).

Despite the decrease in the number of datasets published in 2017, with an index similar to 2014, a return to growth has been observed to date, confirming an upward disposition. In fact, in the current year 2020 the largest number of datasets associated with a virus is recorded. The Covid-19 associated with SARS-CoV-2 presents 94 data sets in just 3 months, which exceeds any forecast.

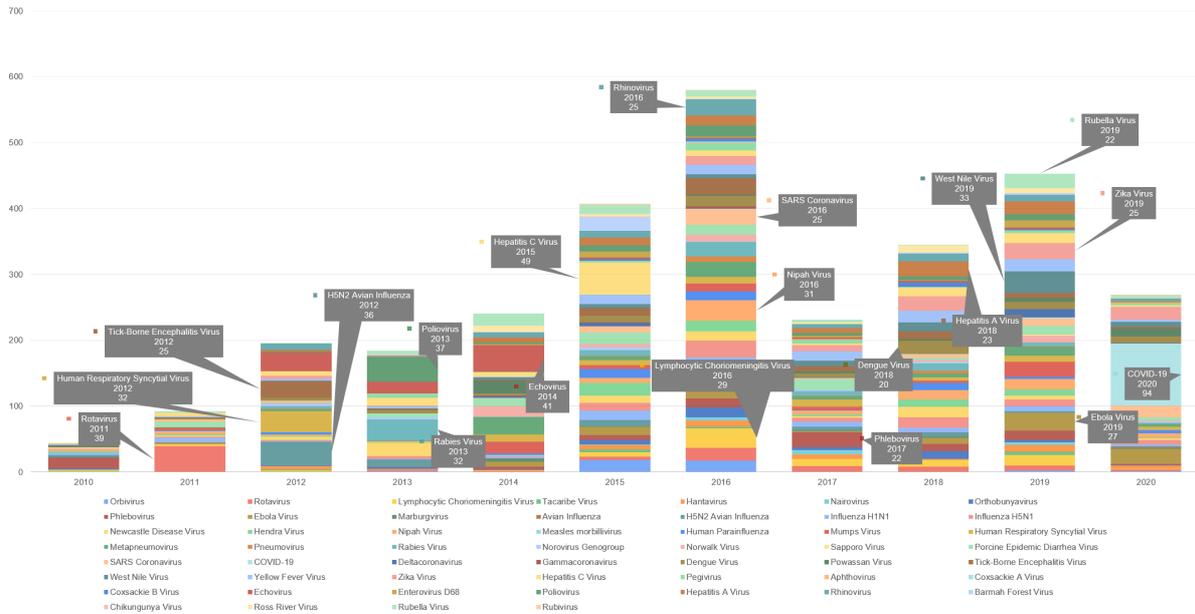

Figure 2. Publication of RNA virus datasets in the last 10 years. Source: Google Dataset Search. Collection date: May 8, 2020. Data available at:
https://github.com/manublaz/datasets/blob/master/cronologiaPublicacionDatasetsVirusARN.xlsx

At this stage of the research, it is worth asking if there is a cause-effect relationship between the main episodes of epidemics and pandemics that have transpired in the scientific literature and the media, in reference to datasets. To answer this question, Figure 3 has been prepared, in which the most well-known epidemics and pandemics are superimposed on the timeline with respect to the publication dates of data sets registered in Google Dataset Search. The results obtained indicate that there is always a delay in the publication of datasets with respect to the dates on which the outbreaks or epidemics are supposed, the most notable case being the SARS-CoV that took place between 2002 and 2004. Yes Well, the first dataset on SARS-CoV is recorded the year after the end of the epidemic, the quantitative values are almost inconsequential until 2012, when most of the data are published, a total of 36 datasets.

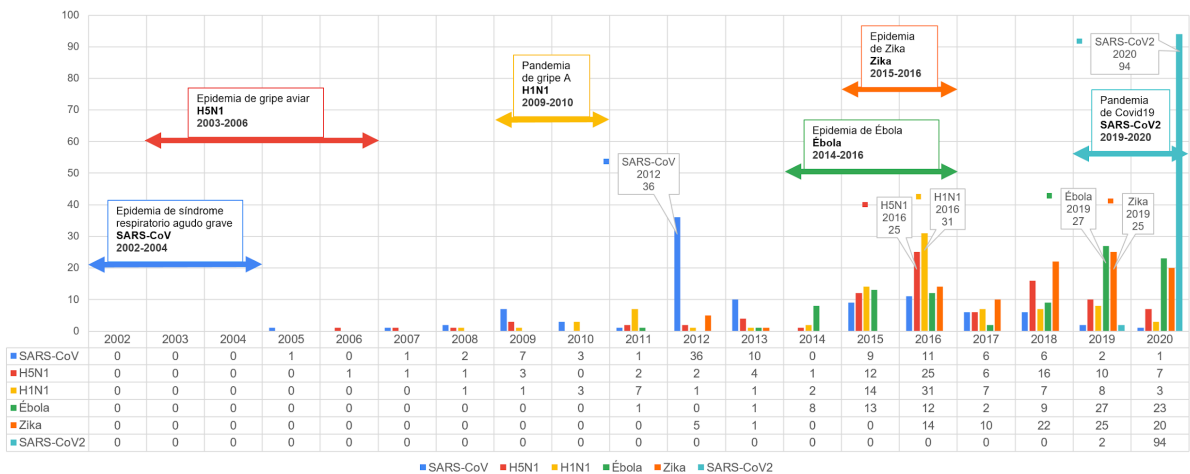

Figure 3. Chronology of epidemics and publication of datasets.
Source: Google Dataset Search. Collection date: May 8, 2020. Data available at
https://github.com/manublaz/datasets/blob/master/cronologiaPublicacionDatasetsPapersEpidemias.xlsx)

It is observed that in 2016 there was an increase in the number of datasets referring to the study of viruses, specifically:

- Lymphocytic Choriomeningitis or Choriomeningitis virus produced by rodents (29 results);
- Nipah virus endemic to the area of Malaysia and India (31 results);
- the Rhinovirus or common cold virus (25 results);
- the SARS Coronavirus or SARS-CoV (25 results).

It is shown that 2016 is one of the most productive in terms of data sets. However, the maximum value is recorded by Covid-19 associated with SARS-CoV-2, which in just three months of 2020 has recorded a total of 94 datasets. Although as indicated above, only 52% of the data is scientific and to a lesser extent, only 15% reusable, it can be said that the severity of the Coronavirus crisis has caused a rapid response from the scientific community that it has had an eloquent reflection in the statistics, although the degree of coordination, aggregation and grouping of the data according to the various producing subjects cannot yet be confirmed.

This pattern is repeated in most cases, for example, in the avian influenza epidemic caused by H5N1 between 2003 and 2006 and influenza A or H1N1, between 2009 and 2010. In both cases, the published dataset figures remain below 10 until 2016, when the largest increase in the historical series occurs. Another notable case is the Ebola epidemic, which occurred between 2014 and 2016, of which datasets published before and during the outbreak were observed. This can easily be explained, since the Ebola virus has been known since 1976 (Emond et al., 1977), since at least seven outbreaks have been recorded. This factor, together with the high mortality and morbidity, have been able to influence the interest of the scientific community. In fact, there is an increase in the number of jobs and the number of published datasets, which reaches its maximum in 2019.

Another similar case is that of the Zika virus. Although datasets have been found since 2012, the Zika epidemic will not take place until 2015 and 2016, when the epidemic outbreak reaches South, Central America and part of the Caribbean. Like Ebola, Zika has been known in advance, since 1947 (Dick; Kitchen; Haddow, 1952), which explains the existence of datasets prior to the last epidemic. However, a growth of datasets has been observed since then, which has its maximum values between 2019 and 2020. If the results obtained for the datasets are compared with the published scientific articles, it is observed that their publication frequency is higher and they overlap with the dates of incidence of pandemics (Figure 4). This phenomenon is clearly justified as it is the preferred means of scientific communication and helps to put the publication of articles and datasets into context, both in volume and periodicity, although its comparative ratio is extremely low.

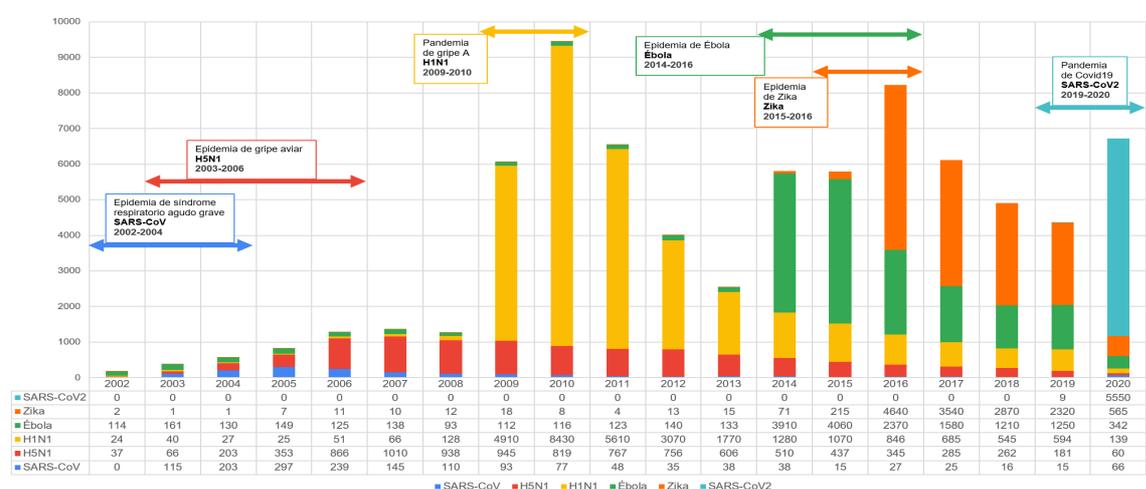

Figure 4. Chronology of epidemics and publication of scientific articles.

Never before has this level of concern been reached, reflected in scientific publications and datasets, which demonstrates the degree of attention of the scientific community to the essential difficulties that affect society as a whole. Other aspects to highlight are the appearance of datasets on SARS-CoV-2, just at the beginning of the Coronavirus crisis in December 2019, as well as the quantitative leap produced in such a short space of time.

The results obtained suggest several reflections on the factors that affect the production of specialized virology datasets, among which the following could be highlighted:

- Severity of the pandemic. It follows that, the more affected and dangerous a virus, the greater the number of published datasets, linked to research and scientific articles. This is confirmed with the SARS-CoV-2 virus, by showing that the rate of publication of datasets per article is the highest compared to the rest of the viruses analyzed.

- Transparency of investigations. A notable percentage of research datasets are not freely accessible, making it difficult to independently verify the scientific facts from which scientific publications are derived (Irwin, 2009). The principles of reproducibility and repeatability are compromised by the interests of the companies that finance research and data mining (Bekelman; MPhil; Gross, 2003).

- Difficulties inherent to the investigation and the integrity of the data. This factor could explain the time lag between a pandemic and the publication of the dataset, as has been described. This may be due to the difficulty in devising adequate scientific methods and trials, validated by the scientific community, the shortcomings in the automation and data collection processes, or the lengthy process of verification and curation of these, typical of this area of research. knowledge (Schneier, 2012).

Finally, to specify the scope of the analyzed information, the providers that collaborate with Google are reviewed, where certain limitations are observed. One of the most relevant is the lack of official figures on its coverage. Therefore, the information on providers analyzed corresponds to the first 100 results that the search engine retrieves, based on the sample of Q1 virus families, Q2 virus species and Q2 sample.

From here it can be stated that the coverage of the datasets and files recovered on RNA viruses varies depending on the provider. Not all data providers provide usable datasets to Google, as indicated; some data providers even show office files, which are of no use to form knowledge bases for big-data. However, from the search engine, no distinction or qualitative assessment is made in this regard.

As shown in table 4, two data providers stand out above the rest: Figshare and Researchgate. Regarding the general search for virus families (Q1), both have more than 70% coverage and the specific searches for each family (Q2) are close to 60%. The rest of the analyzed providers (Statista, Omicsdi, Datamed, Search.datacite, etc.) have very low coverage, individually below 4%, but globally they accumulate slightly less than 30% in Q1 and 40% in Q2, approximately.

Table 4. Main data providers of the analyzed datasets

| Data providers | Q1 | | Q2 | | Q2 Sample | |
|---|---|---|---|---|---|---|
| Figshare, Plos.figshare, Springernature.figshare.com | 658 | 54,96% | 1183 | 33,45% | 133 | 51,55% |
| Researchgate | 231 | 16,75% | 877 | 24,8% | 47 | 14,20% |
| Catalog.data.gov | 23 | 1,66% | 64 | 1,8% | 12 | 4,65% |
| Statista | 49 | 3,55% | 194 | 5,48% | 12 | 4,65% |
| *Omicsdi* | 126 | 0,58% | 294 | 8,31% | 12 | 4,65? |
| *Datamed* | 44 | 3,19% | 133 | 3,76% | 1 | 0,38% |
| *Search.datacite* | 42 | 3,04% | 225 | 6,36% | 1 | 0,38% |
| Other data providers | 106 | 7,6% | 566 | 16% | 44 | 43,50% |
| **n. total datasets retrieved** | **1.379** | **100%** | **3.536** | **100%** | **258** | **100%** |

As mentioned in the analysis of the results, 37% of the scientific datasets are not reusable or accessible. This is also due in part to the fact that providers such as Researchgate and Statista in many cases limit the data displayed, requiring registration, subscription or prior payment, for full access. This type of practice is far from the concept of open and collaborative science, opening up the reflection on the transfer of dataset rights in the main data providers and repositories.

Another relevant issue is the adequacy of the Schema.org metadata to the description of the datasets. It has been observed that Google Dataset Search collects your information from different data providers. It should be noted that the Schema.org metadata is a very versatile format, since it adopts the main description matters for this type of document. For example, the dataset's membership relationship to other data collections, summary, authorship and its affiliation, citations, comments, access conditions, edition status, creation and update dates, editions, encoding, resource link, distribution, spatial and temporal coverage, among others. Although it is a very complete model, if the description of each section is taken care of, there are two fundamental aspects, which have not yet been adequately addressed:

- the control of the versions of the datasets, in order to track the changes produced and facilitate their recovery. In this sense, the Schema.org metadata should include a tuple in which the dates of the dataset versions are associated with their link to the original file and the statement of responsibility, similar to the operation already known in software repositories such as Github ( Blischak; Davenport; Wilson, 2016);
- the lack of a definition of the dataset's data structures, which makes it difficult to unify datasets with similar purposes and themes. That is, the introduction of a field that defines, as a comma-separated list of values, the field header of the dataset in question. This makes it easy to quickly identify and compare the collection of datasets, to discover which ones can be grouped together to generate large collections of data. Therefore, it also simplifies the automation of correspondence mapping, between different dataset fields, for their merging, if necessary.

The normalization of metadata seeks to guarantee consistency in the format of the datasets retrieved by Google Dataset Search and to be able to offer users a meaningful and unified search experience (Canino, 2019). For example, that the updates of the epidemiological situation of Covid-19, are not broken down into provinces and update dates, but can be studied in a single dataset that integrates them all, maintaining the identification of their origin, date and version. This type of case is frequent among search engine results.

Regarding the reuse of datasets, it can be said that their dissemination and use is its main purpose, since in most cases it involves the registration of scientific tests, experiments and observations. This gives the dataset the value of scientific evidence on which the conclusions and results of many scientific papers are based. The nature of this type of document makes it a valuable resource or source of information, as long as its correct identification and aggregation is possible. In this task, there is data and text mining, and specifically scraping techniques (Singhal; Srivastava, 2013), which make the automatic extraction of datasets possible, from different data providers, or, as This is the case of a specialized search engine. This allows you to focus your search on a specific topic or issue, to obtain the most relevant datasets and process them in one or more tables of a database. Obviously, there is not yet a fully automated method by which all fields in a dataset can be automatically mapped or mapped. Hence the importance of having an adequate metadata system, which makes up for this problem, as has been explained.

Subsequently, the information from the datasets can be enriched by other sources of information, to compose a more complete collection with which to obtain a new scientific analysis, in what has been called the scientific data curation process (Karasti; Baker; Halkola, 2006). The variety of datasets around an object of study, as well as their volume, favor their reuse in big data analysis. This is so because the purpose is the correlation of data sets from various sources. For example, the relationship between the compounds of the drugs used in the treatment of Covid-19 and the evolution of the patients, the medical records and their hematological tests (Wang; Ng; Brook, 2020). They constitute differentiated data sets that are indexed, grouped and related, to infer a cause and effect relationship or a Pareto cumulative probability distribution that facilitates the patient's diagnosis (Ahlawat; Chug; Singh, 2019). This implies the multidimensional storage of the data sets (Elmeiligy; ElDesouky; Elghamrawy, 2020), where each dimension is an analysis factor, which is decomposed for its identification and classification in nodes, also called pairs sets, to be correlated. with other nodes from other datasets. This is possible thanks to Map-reduce techniques, capable of quantifying the frequency of appearance of the elements of each node, resulting in a combined and ordered value that reflects the weight of each relationship (Khashan et al., 2020). These types of results help to determine which factors are decisive in the improvement or worsening of a disease, promoting the subsequent development of machine learning models.

## 4. Conclusions

1. This work shows that not all the results displayed by the search engine are datasets. Although there is confusion with the term, it must be emphasized that a dataset or dataset is that collection of data structured and delimited in its values, so that it can be reused in databases, spreadsheets, statistical analysis programs and big data. The dataset's own formats are CSV separated by comma values, SQL because it constitutes the structured query language of databases and derived from XML, because it allows the precision of values through labels and extensible markup structures. In fact, according to (Qian; Bailey; Leckie, 2006) it is that collection of objects or data, represented successively following a pattern or tabulation scheme, which facilitates their instantiation and compilation.

2. The analysis shows that approximately 15% of the results are reusable datasets of proven scientific provenance, which represents a reduced amount of the information available in the search engine. Although open science is consolidating at the level of scientific articles (Mckiernan et al., 2016), it is not doing so in the field of data sets. This conclusion may be due to the lack of supervision, filtering and evaluation of the datasets, after indexing in the search engine. According to Google, the information in the datasets is collected directly through the Schema.org metadata. Perhaps this procedure does not trigger a true review of the type of data, format, origin, quality and reliability of the same, resulting in an automatic registration of the information. Clearly, Google's platform still needs to improve its data entry and verification process if it is to become a reliable source of scientific data. Other aspects that could be improved are the limitation of results to only a hundred per search, the lack of the filtering system to correctly distinguish the download formats according to their extensions, the scientific or commercial origin of the dataset, country or region of collection, language, dates. of publication and update, officiality of the data, affiliation, origin, secondary producers or collaborators, provider or data aggregator and filtering by date ranges.

3. Google Dataset Search is shown as a singular aggregator, which obtains information from multiple data providers, among which there is a notable quantitative inequality, at least in the results obtained for families of RNA viruses. The main data provider is Figshare. In the general search for virus families (Q1), 54.96% of information was recovered, in the specific searches for each family (Q2) 33.45% and in the specific sample of Q2, 51.55%. In contrast, other providers such as Catalog.data.gov, Statista, Omicsdi, Datamed and Search.datacite, do not exceed 9% in each of the three searches performed.

4. In order to increase the percentage of reusable datasets, Google should collect among its selection criteria, screening or filtering the datasets whose formats are CSV, XLS, SQL and XML and establish a clear differentiation between commercial, scientific, and article datasets scientific and other products of scientific publications, such as figures, illustrations and presentations. It has been found that there is a great difference between the data that the search engine returns, compared to the real data. In fact, only 52% of the datasets are of scientific origin, with 15% being really reusable. 37% of the results were badly cataloged, considering datasets to the illustrations, figures and presentations of scientific publications.

5. The research seems to confirm challenges that still need to be overcome, both by data providers and dataset finders. Specific: a) The standardization and improvement of the metadata format for the exchange of datasets, so that the versions and additions of the authors and institutions participating in the edition can be distinguished. b)The full-text indexing of the datasets, including the list of fields, figures and registered character strings, so that the retrieval does not depend exclusively on the description provided in the metadata. c)Show the relationships between scientific articles and the datasets on which they are based, in order to study their impact on scientific advances. d) Classify the datasets according to their scientific discipline, applications, temporal and geographic coverage. e) Develop big data techniques to detect patterns of similarity and correlation between datasets, in order to facilitate the researcher in selecting the appropriate data sets.

6. Regarding the analysis of the results, a notable increase in the publication of virus datasets has been observed since 2016, which temporarily coincides with the Ebola and Zika epidemics. Among the RNA viruses with the most datasets, the Choriomeningitis virus, the Nipah virus, the Rhinovirus, the SARS-CoV and recently SARS-CoV-2 stand out. In fact, they have exceeded the dataset publication rate in a period of just 3 months. None of the analyzed viruses reached these levels, which shows an evident reaction from the scientific community. Thus, it can be stated that the mortality and morbidity of a virus in the population are factors that are intrinsically related to the number of published datasets and papers. This is demonstrated by comparing the chronologies of the main epidemics, being especially true in the case of H1N1, Ebola, Zika and SARS-CoV-2. It can also be concluded that the severity of a pandemic, transparency in research, as well as difficulties in collecting data and designing scientific methods for the development of clinical trials, may be some of the causes of a low frequency of dataset publication, especially if it is compared with other documentary types, such as scientific papers. This imbalance is widened when the data set is expected to be open access in its completion, being one more barrier that the open science scheme must overcome.

## 5. Future lines of research

This work demonstrates the importance of datasets in the context of Information and Documentation, since they constitute a source of fundamental knowledge for scientific production. Data collections are the primary documentary type that collects the observations of a scientific experiment, or its indicators and evaluation factors. Therefore, in general, any research that addresses its better knowledge, administration, recovery and processing, will facilitate the advancement and dissemination of Science. It should not be forgotten that Documentation, as an auxiliary science, requires an open knowledge, both of the information needs, and of the documentary types that are demanded and used. Therefore, there is room for future related research, for example:

1. Analysis of DNA virus datasets and their comparison with the results of RNA viruses. By replicating the same research method, DNA-type viruses can be taken as the basis of consultation, in order to be compared with the results obtained in this research. In this way, it would be possible to confirm whether this other type of virus has a different publication rate, as well as its incidence in the publication of scientific articles, its dependence on external factors such as epidemics or outbreaks, or the presence of differentiating elements with respect to to another data set. The results obtained can provide the necessary experience for the design of better search and aggregation applications for this type of data sets in the area of knowledge of Virology.

2. Replication of the dataset research method in other knowledge or subject areas. The method used is applicable to any topic. This is so, since the section of the thesaurus for consultation can be varied, or it can be replaced by a controlled vocabulary, standardized and recognized by the scientific community. This allows targeted queries, in accordance with official terminology, in the dataset search engine and obtain comparable results. This type of study would allow to know which areas of knowledge generate more data collections, for what reasons, under what conditions, as well as their correlation with the main scientific publications.

3. Study of the correlation between scientific publications, the use of datasets and their value. It can be affirmed that scientific investigations supported by datasets, coming from experimentation, observation and data collection, are in better conditions for the demonstration and empirical justification, than those that do not enjoy said foundations. Although this hypothesis is reasonable, it would be expected to be confirmed by analyzing a sample of articles, based on scientific datasets and open access, in which their weight in obtaining citations would be analyzed. That is, to determine the degree of influence that this aspect supposes, for the success of a scientific article.

4. Design of a metadata format suitable for scientific datasets. Although the Schema.org metadata format is the current reference, it seems logical to propose an improvement or update of its fields, to adapt it to version control and the identification of the data structures of the datasets, to favor their aggregation. This research would help to improve the retrieval of this type of document in search engines such as Google Dataset Search, providing the opportunity to simplify the researcher's task.

## 6.References

**Ahlawat, Khyati**; **Chug, Anuradha**; **Singh, Amit-Prakash** (2019). "Empirical evaluation of map reduce based hybrid approach for problem of imbalanced classification in big data". *International journal of grid and high performance computing*, v. 11, n. 3, pp. 23-45.
*https://doi.org/10.4018/IJGHPC.2019070102*

**Bekelman, Justin E.**; **MPhil, Yan-Li**; **Gross, Cary P.** (2003). "Scope and impact of financial conflicts of interest in biomedical research: a systematic review". *Jama*, v. 289, n. 4, pp. 454-465.
*https://doi.org/10.1001/jama.289.4.454*

**Blischak, John D.**; **Davenport, Emily R.**; **Wilson, Greg** (2016). "A quick introduction to version control with Git and GitHub". *PLoS computational biology*, v. 12, n. 1.
*https://doi.org/10.1371/journal.pcbi.1004668*

**Brickley, Dan**; **Burgess, Matthew**; **Noy, Natasha** (2019). "Google Dataset Search: Building a search engine for datasets in an open web ecosystem". In: *Proceedings of the world wide web conference*, pp. 1365-1375.
*https://doi.org/10.1145/3308558.3313685*

**Broder, Andrei** (2002). "A taxonomy of web search". *ACM Sigir forum*. v. 36, n. 2, pp. 3-10.
*https://doi.org/10.1145/792550.792552*

**Canino, Adrienne** (2019). "Deconstructing Google Dataset Search". *Public services quarterly*, v. 15, n. 3, pp. 248-255.
*https://doi.org/10.1080/15228959.2019.1621793*

**Chen, Emily**; **Lerman, Kristina**; **Ferrara, Emilio** (2020). "Tracking Social Media Discourse About the COVID-19 Pandemic: Development of a Public Coronavirus Twitter Data Set". *JMIR Public health and surveillance*, v. 6, n. 2.
*https://arxiv.org/abs/2003.07372*
*https://doi.org/10.2196/19273*

**Chen, Serena H.**; **Young, M. Todd**; **Gounley, John**; **Stanley, Christoher**; **Bhowmik, Debsindhu** (2020). "Distinct structural flexibility within SARS-CoV-2 spike protein reveals potential therapeutic targets". *BioRxiv*.
*https://doi.org/10.1101/2020.04.17.047548*

**Corrales-Garay, Diego**; **Ortiz-de-Urbina-Criado, Marta**; **Mora-Valentín, Eva-María** (2019). "Knowledge areas, themes and future research on open data: A co-word analysis". *Government information quarterly*, v. 36, n. 1, pp. 77-87.
*https://doi.org/10.1016/j.giq.2018.10.008*

**Dick, George W. A.**; **Kitchen, Stuart F.**; **Haddow, Alexander J.** (1952). "Zika virus (I). Isolations and serological specificity". *Transactions of the royal society of tropical medicine and hygiene*, v. 46, n. 5, pp. 509-520.
*https://doi.org/10.1016/0035-9203(52)90042-4*

**Elmeiligy, Manar A.**; **ElDesouky, Ali I.**; **Elghamrawy, Sally M.** (2020). "A multi-dimensional big data storing system for generated Covid-19 large-scale data using Apache spark". *arXiv preprint.*

*https://arxiv.org/abs/2005.05036*

**Emond, Ronald T.**; **Evans, Barry**; **Bowen, Ernest-Thomas**; **Lloyd, Graham** (1977). "A case of Ebola virus infection". *British medical journal*, v. 2, n. 6086, pp. 541-544.
*https://doi.org/10.1136/bmj.2.6086.541*

*Google Search* (2020). *dataset*.
*https://developers.google.com/search/docs/data-types/dataset*

**Haleem, Abid**; **Javaid, Mohd**; **Khan, Ibrahim-Haleem**; **Vaishya, Raju** (2020). "Significant applications of big data in Covid-19 Pandemic". *Indian journal of Orthopaedics*, v. 54, n. 7.
*https://doi.org/10.1007/s43465-020-00129-z*

**Hawking, David**; **Craswell, Nick**; **Bailey, Peter**; **Griffihs, Kathleen.** (2001). "Measuring search engine quality". *Information retrieval*, v. 4, n. 1, pp. 33-59.
*https://doi.org/10.1023/A:1011468107287*

**Hawking, David**; **Craswell, Nick**; **Thistlewaite, Paul**; **Harman, Dona** (1999). "Results and challenges in web search evaluation". *Computer networks*, v. 31, n. 11-16, pp. 1321-1330.
*https://doi.org/10.1016/S1389-1286(99)00024-9*

**Hernández-Pérez, Tony** (2016). "En la era de la web de los datos: primero datos abiertos, después datos masivos". *El profesional de la información*, v. 25, n. 4, pp. 517-525.
*https://doi.org/10.3145/epi.2016.jul.01*

**Howe, Nicola**; **Giles, Emma**; **Newbury-Birch, Dorothy**; **McColl, Elaine** (2018). "Systematic review of participants' attitudes towards data sharing: a thematic synthesis". *Journal of health services research & policy*, v. 23, n. 2, pp. 123-133.
*https://doi.org/10.1177/1355819617751555*

**Irwin, Richard S.** (2009). "The role of conflict of interest in reporting of scientific information". *Chest*, v. 136, n. 1, pp. 253-259.
*https://doi.org/10.1378/chest.09-0890*

**Johansson, Michael A.**; **Saderi, Daniela** (2020). *"Open peer-review platform for COVID-19 preprints"*. *Nature*, v. 579, n. 7797.
*https://doi.org/10.1038/d41586-020-00613-4*

**Karasti, Helena**; **Baker, Karen S.**; **Halkola, Eija** (2006). "Enriching the notion of data curation in e-science: data managing and information infrastructuring in the long term ecological research (LTER) network". *Computer supported cooperative work*, v. 15, n. 4, pp. 321-358.
*https://doi.org/10.1007/s10606-006-9023-2*

**Khashan, Eman A.**; **ElDesouky, Ali I.**; **Fadel, Magdy**; **Elghamrawy, Sally M.** (2020). "A big data based framework for executing complex query over Covid-19 datasets (Covid-QF)". *arXiv preprint arXiv:2005.12271*.
*https://arxiv.org/abs/2005.12271*

**King, John-Douglas**; **Li, Yuefeng**; **Tao, Xiaohui**; **Nayak, Richi** (2007). "Mining world knowledge for analysis of search engine content". *Web intelligence and agent systems: An international journal*, v. 5, n. 3, pp. 233-253.

*https://dl.acm.org/doi/10.5555/1377776.1377777*

**Landau, Yuval; Kiryati, Nahum** (2019). "Dataset growth in medical image analysis research". arXiv arXiv:1908.07765.
*https://arxiv.org/abs/1908.07765*

**Le-Guillou, Ian** (2020). "Covid-19: How unprecedented data sharing has led to faster-than-ever outbreak research". *Horizon. The UE research & innovation magazine*, 23 March.
*https://horizon-magazine.eu/article/covid-19-how-unprecedented-data-sharing-has-led-faster-ever-outbreak-research.html*

**Lewandowski, Dirk** (2015). "Evaluating the retrieval effectiveness of web search engines using a representative query sample". *Journal of the Association for Information Science and Technology*, v. 66, n. 9, pp. 1763-1775.
*https://asistdl.onlinelibrary.wiley.com/doi/10.1002/asi.23304*
*https://doi.org/10.1002/asi.23304*

**López-Borrull, Alexandre**; **Ollé-Castellà, Candela**; **García-Grimau, Francesc**; **Abadal, Ernest** (2020). "Plan S y ecosistema de revistas españolas de ciencias sociales hacia el acceso abierto: amenazas y oportunidades". *El profesional de la información*, v. 29, n. 2.
*https://doi.org/10.3145/epi.2020.mar.14*

**Marcial, Laura-Haak**; **Hemminger, Bradley M.** (2010). "Scientific data repositories on the Web: An initial survey". *Journal of the American Society for Information Science and Technology*, v. 61, n. 10, pp. 2029-2048.
*https://doi.org/10.1002/asi.21339*

**McKiernan, Erin C.**; **Bourne, Philip E.**; **Brown, C. Titus**; **Buck, Stuart**; **Kenall, Amye**; **Lin, Jennifer**; **McDougall, Damon**; **Nosek, Brian A.**; **Ram, Karthik**; **Soderberg, Courtney K.**; **Spies, Jeffrey R.**; **Thaney, Kaitlin**; **Updegrove, Andrew**; **Woo, Kara H.**; **Yarkoni, Tal** (2016). "Point of view: How open science helps researchers succeed". *Elife*, v. 5, e16800.
*https://doi.org/10.7554/eLife.16800.001*

**Mello, Michelle M.**; **Lieou, Van**; **Goodman, Steven N.** (2018). "Clinical trial participants' views of the risks and benefits of data sharing". *New England journal of medicine*, v. 378, n. 23, pp. 2202-2211.
*https://doi.org/10.1056/NEJMsa1713258*

**Nosek, Brian A.**; **Alter, George**; **Banks, George C.**; **Borsboom, Denny**; **Bowman, Sara D.**; **Breckler, Steven J.**; **Buck, Stuart**; **Chambers, Christopher D.**; **Chin, Gilbert**; **Christensen, Garret**; **Contestabile, M.**; **Dafoe, A.**; **Eich, Eric**; **Freese, J.**; **Glennerster, R.**; **Goroff, D.**; **Green, Donald P.**; **Hesse, Bradford W.**; **Humphreys, M.**; **Ishiyama, John**; **Karlan, D.**; **Kraut, A.**; **Lupia, A.**; **Mabry, Patricia L.**; **Madon, T.**; **Malhotra, N.**; **Mayo-Wilson, Evan**; **McNutt, M.**; **Miguel, Edward**; **Levy-Paluch, Elizabeth**; **Simonsohn, U.**; **Soderberg, Courtney**; **Spellman, Barbara A.**; **Turitto, J.**; **VandenBos, Gary-Roger**; **Vazire, Simine**; **Wagenmakers, E. J.**; **Wilson, R.**; **Yarkoni, T.** (2015). "Promoting an open research culture". *Science*, v. 348, n. 6242, p.p. 1422–1425.
*https://doi.org/10.1126/science.aab2374*

**Polonetsky, Jules**; **Tene, Omer**; **Finch, Kelsey** (2016). "Shades of gray: Seeing the full spectrum of practical data de-intentification". *Santa Clara law review.* v. 56, n. 593, pp. 593-618.
*https://digitalcommons.law.scu.edu/cgi/viewcontent.cgi?article=2827&context=lawreview*

**Qian, Xiaoyuan**; **Bailey, James**; **Leckie, Christopher** (2006). "Mining generalised emerging patterns". En: Sattar, Abdul; Kang, Byeong-Ho (eds.). *Australasian joint conference on artificial intelligence*. Berlin, Heidelberg: Springer, pp. 295-304. ISBN: 978 3 540 49788 2
*https://doi.org/10.1007/11941439_33*

**Saheb, Tahereh; Izadi, Leila** (2019). "Paradigm of IoT big data analytics in healthcare industry: a review of scientific literature and mapping of research trends". *Telematics and informatics*, v. 41, pp. 70-85
*https://doi.org/10.1016/j.tele.2019.03.005*

**Schneier, Bruce** (2012). "Securing medical research: A cybersecurity point of view". *Science*, v. 336, n. 6088, pp. 1527-1529.
*https://doi.org/10.1126/science.1224321*

Science Europe (2019). *Plan S: Making full and immediate Open Access a reality.*
*https://www.scienceeurope.org/coalition-s/*

**Singhal, Ayush**; **Srivastava, Jaideep** (2013). "Data extract: Mining context from the web for dataset extraction". *International journal of machine learning and computing,* v. 3, n. 2, pp. 219-223.
*https://doi.org/10.7763/IJMLC.2013.V3.306*

**Wang, C. Jason**; **Ng, Chun Y.**; **Brook, Robert H.** (2020). "Response to Covid-19 in Taiwan: big data analytics, new technology, and proactive testing". *Jama*, v. 323, n. 14, pp. 1341-1342.
*https://doi.org/10.1001/jama.2020.3151*

**Weston, Sara J.**; **Ritchie, Stuart J.**; **Rohrer, Julia M.**; **Przybylski, Andrew K.** (2019). "Recommendations for increasing the transparency of analysis of preexisting data sets". A*dvances in methods and practices in psychological science,* v. 2, n.3, pp. 214-227.
*https://doi.org/10.1177/2515245919848684*

**Zhou, Chenghu**; **Su, Fenzhen**; **Pei, Tao**; **Zhang, An**; **Du, Yunyan**; **Luo, Bin**; **Cao, Zhidong**; **Wang, Juanle**; **Yuan, Wen**; **Zhu, Yunqiang**; **Song, Ci**; **Chen, Jie**; **Xu, Jun**; **Li, Fujia**; **Ma, Ting**; **Jiang, Lili**; **Yan, Fengqin**; **Yi, Jiawei**; **Hu, Yunfeng**; **Liao, Yilan**; **Xiao, Han** (2020). "Covid-19: challenges to GIS with big data". *Geography and sustainability*, v. 1, n, 1, pp. 77-87.
*https://doi.org/10.1016/j.geosus.2020.03.005*